\documentclass[aoas,preprint,english]{imsart}

\RequirePackage[OT1]{fontenc}
\RequirePackage{amsthm,amsmath}
\RequirePackage[authoryear]{natbib}
\RequirePackage[colorlinks,citecolor=blue,urlcolor=blue]{hyperref}

\RequirePackage{multirow}
\RequirePackage{graphicx}
\RequirePackage[labelformat=empty]{subfig}
\RequirePackage{mathtools}
\RequirePackage{babel}
\RequirePackage{mathrsfs}
\RequirePackage{dsfont}
\startlocaldefs
\numberwithin{equation}{section}
\theoremstyle{plain}

\endlocaldefs

\begin{document}

\begin{frontmatter}
\title{Logistic Biplot for Nominal Data}
\runtitle{Logistic Biplot for Nominal Data}

\begin{aug}
\author{\snm{Julio C\'esar Hern\'andez S\'anchez}\ead[label=jchs]{juliocesar.hernandez.sanchez@ine.es}}
\and
\author{\snm{Jos\'e Luis Vicente-Villard\'on}
\ead[label=jlvv]{villardon@usal.es}}

\affiliation{Spanish Statistical Office and University of Salamanca}

\address{Julio C\'esar Hern\'andez S\'anchez\\
Spanish Statistical Office\\
Zamora\\
Spain\\
Jos\'e Luis Vicente-Villard\'on\\
Salamanca University\\
Salamanca\\
Spain\\
\printead{jchs}\\
\phantom{E-mail:\ }\printead*{jlvv}}

\end{aug}

\begin{abstract}

Classical Biplot Methods allow for the simultaneous representation of individuals (rows) and variables (columns) of a data matrix. For Binary data, Logistic biplots have been recently developed. When data are nominal, linear or even binary logistic biplots are not adequate and techniques as Multiple Correspondence Analysis (MCA), Latent Trait Analysis (LTA) or Item Response Theory for nominal items should be used instead. 

In this paper we extend the binary logistic biplot to nominal data. The resulting method is termed Nominal Logistic Biplot, although the variables are represented as convex prediction regions rather than vectors. Using the methods from Computational Geometry, the set of prediction regions is converted to a set of points in such a way that the prediction for each individual is established by its closest ``category point''. 

Then interpretation is based on distances rather than on projections. We study the geometry of such a representation and construct computational algorithms for the estimation of parameters and the calculation of prediction regions. Nominal Logistic Biplots extend both MCA and LTA in the sense that gives a graphical representation for LTA similar to the one obtained in MCA.
\end{abstract}


\end{frontmatter}

\section{Introduction}

The biplot method \citep{Gabriel71} is a simultaneous graphical representation of the rows and columns of a data matrix.  In practice, biplot fitting occurs either by computing the singular value decomposition (SVD) of the data matrix or by performing an alternating regressions procedure \citep{GabrielZamir1979}. \cite{Jongman-87} fit the biplot by alternating a regression and a calibration step, essentially equivalent to the alternating regressions. \cite{Gower-96} use the term interpolation rather than calibration.
For data with distributions from the exponential family, \cite{Gabriel1998}, describes ``bilinear regression'' as a method to estimate biplot parameters, but the procedure have never been implemented and the geometrical properties of the resulting representations have never been studied.    Method is called ``External Logistic Biplot''. \cite{deLeeuw06} proposes Principal Components Analysis for Binary data based on an alternate procedure in which each iteration is performed using iterative majorization and \cite{Lee10} extends the procedure for sparse data matrices, none of those describe a biplot representation for binary data. \cite{Vicente-06} propose a biplot representation based on logistic responses called ``Logistic Biplot'' that is linear, the paper studies the geometry of this kind of biplots and uses a estimation procedure that is slightly different from Gabriel's method. A heuristic version of the procedure for large data matrices in which scores for individuals are calculated with an external procedure as Principal Components Analysis is described in \cite{Demeyetal}. 

When data are nominal, there are many techniques to deal with it, some of them see the problem from a Factor Analytic point of view to obtain latent factors that explain the correlation among variables, others as some kind of non-parametric approximations to explore the similarities among individuals (Principal Coordinates Analysis or Multidimensional Scaling) but there is a lack of general exploratory techniques for the simultaneous representation of individuals and variables except Multiple Correspondence Analysis, based on the chi-squared distance, that is not always adequate to describe similarities among individuals and correlations among variables. As we will see, it is possible to combine the Factor Analytic approach with the exploratory point of view to obtain a simultaneous representation of individuals and variables (Biplot) that helps to explore the information provided by the data. In this paper we propose ``Nominal Logistic Biplots'' that share characteristics from the previously mentioned techniques; on the one hand is a procedure for dimension reduction, explaining the correlation among nominal variables with a reduced number of latent factors and on the other hand can serve as a exploratory biplot technique. 
Nominal Logistic Biplots represent the rows of a data matrix as points on a reduced dimension representation (usually 2 or 3) and variables as prediction regions (convex polygons), in the same way as is done in \cite{Gower-96}  for Multiple Correspondence Analysis. For MCA the category points are calculated first and then the prediction regions are obtained as regions of a voronoi diagram; in this case the prediction regions are obtained first by nominal logistic regression that defines tessellations of the space; the problem is then finding the voronoi diagram that is the closest to the ``logistic tessellation'' and a set of generators for such a diagram are the category points, the main advantage of doing so is that the interpretation of the biplot is done in terms of distances, for each individual the predicted category is the closest to it on the biplot.

There are several candidate methods for parameter estimation:
\begin{itemize}
\item \textbf{Alternated generalized regressions and interpolations.} (Joint Maximum Likelihood, \citep{Gabriel1998,Vicente-06}).
\item \textbf{Marginal Maximum Likelihood} (As in Item Response Theory, \citep{Baker92,Bock81,Chalmers2012}).
\item \textbf{External Logistic Biplots}: Heuristic approach for big data matrices. (Logistic fits on the Principal Coordinates, \citep{Demeyetal}).
\end{itemize}

In the context of binary logistic biplots the first two procedures are particularly useful when the number of individuals (companies) is higher than the number of variables (indicators), being the second more stable for cases with a high number of individuals. The third method is more useful when the number of indicators is higher than the number of companies, although it can be applied in any case. 
In this paper we have chosen a version of the second method. Final estimation of the variable parameters where calculated using an algorithm developed for this paper, standard logistic regressions on the scores provided by mirt or by Principal Coordinates Analysis. 

In Section 2 we describe linear and logistic biplots as a basis for the development of the nominal case. Section 3 describes the model and its main geometrical characteristics. Section 4 presents an algorithm to obtain the category points for each variable. Section 5 applies the nominal logistic biplot to a classical set of data and Section 6 concludes the paper with a discussion and some suggestions for further research.

\section{Linear and Binary Logistic Biplots}
\subsection{Classical Linear Biplots and the Singular Value Decomposition}

Let ${{\bf{X}}_{I \times J}}$  be a data matrix containing the measures of $J$ variables ( continuous) on $I$ individuals. A S-dimensional biplot is a graphical representation of a data matrix  $\bf{X}$ by means of markers (points or vectors) ${{\bf{a}}_1}, \ldots ,{{\bf{a}}_I}$ for its rows and markers ${{\bf{b}}_1}, \ldots ,{{\bf{b}}_J}$ for its columns, in such a way that the product ${{\bf{a'}}_i}{{\bf{b}}_j}$ approximates the element $x_{ij}$ as close as possible. Arranging the markers as row vectors in two matrices $\bf{A}$ and $\bf{B}$, the approximation of $\bf{X}$ can be written as ${\bf{X}} \approx {\bf{AB'}}$.  Although the classical biplot is well known, we include here a short description, in terms of alternating regressions, related to our proposal.

The most typical way to obtain the biplot is from the singular value decomposition.
Let $R=rank(\bf{X})$, there exists a factorization of the form
\begin{equation}
\label{eqSVD}
{\bf{X}} = {\bf{U}}\Lambda {\bf{V'}} = \sum\limits_{r = 1}^R {{\lambda _r}} {{\bf{u}}_r}{{\bf{v'}}_r}
\end{equation}
where $\bf{U}$ is an ${I \times R}$ unitary matrix, $\bf{\Lambda}$ is an ${R \times R}$ diagonal matrix with non-negative real numbers on the diagonal, and $\bf{V}$ an ${J \times R}$ unitary matrix. Such a factorization is called the singular value decomposition of $\bf{X}$. The diagonal entries $\lambda_{r,r}$ of $\bf{\Lambda}$ are known as the singular values of $\bf{X}$, and are placed in decreasing order, and the columns ${{\bf{u}}_r}$ and ${{\bf{v}}_r}$ of $\bf{U}$ and $\bf{V}$ are known as left and right singular vectors. The \emph{Singular Value Decompositions} are closely related to the \emph{Eigen Decompositions},  the columns of $\bf{U}$ are the eigenvectors of ${\bf{XX'}}$, the columns $\bf{V}$ the eigenvectors of ${\bf{X'X}}$ and the diagonal elements of $\bf{\Lambda}$ are the squared roots of the non-null eigenvalues of both matrices (that are the same).  

It is known that the best $S-rank$ approximation of $\bf{X}$ is given by its first \emph{S} singular values and vectors
\begin{equation}
\label{eqSVD2}
{\bf{X}} \cong \sum\limits_{s = 1}^S {{\lambda _s}{{\bf{u}}_s}{{{\bf{v'}}}_s}}  = {{\bf{U}}_{(S)}}{\Lambda _{(S)}}{{{\bf{V'}}}_{(S)}}
\end{equation}

From he SVD it is easy to obtain a factorization in the Biplot form with the desired restriction taking
\begin{equation}
\label{eqgamma3}
{\bf{A}} = {{\bf{U}}_{(S)}}\Lambda _{(S)}^\gamma ,\quad {\bf{B}} = {\bf{V}}_{(S)}^{1 - \gamma }
\end{equation}
with $0 \le \gamma  \le 1$, as row and column coordinates respectively. This will be referred in the later as PCA-Biplot or Classical Biplot. For example, with $\gamma = 1$, $\bf{A}$ are the coordinates of individuals on the principal components and $\bf{B}$ are the eigenvectors of the covariance matrix.

There is another way of obtaining biplots from alternated regressions. If we consider the row markers $\bf{A}$, as fixed, the column markers can be computed by regression 
\begin{equation}
{\bf{B'}} = {({\bf{A'A}})^{ - 1}}{\bf{A'X}}
\label{reg1}
\end{equation}
In the same way, fixing $\bf{B}$, $\bf{A}$ can be obtained as
\begin{equation}
{\bf{A'}} = {({\bf{B'B}})^{ - 1}}{\bf{B'X'}}
\label{reg2}
\end{equation}
Alternating the steps (\ref{reg1}) and (\ref{reg2}) the product converges to the SVD. The algorithm can then be completed with an orthogonalization step to ensure the uniqueness of its solution. The regressions in (\ref{eqSVD}) and (\ref{eqSVD2}) can be separated for each row and column of the data matrix. This symmetrical process is commonly used to adjust bilinear (or bi-additive) models with symmetrical roles for rows and columns.  For a data matrix of individuals by variables, the roles of rows and columns are non-symmetrical, nevertheless the algorithm is still valid and is interpreted as a two-step process, alternating a regression step and an interpolation/calibration step.
The regression step adjusts a separate linear regression for each column (variable) and the interpolation step interpolates an individual using the column markers as the reference. Geometry of the interpolation step is described in \cite{Gower-96}.

\subsection{Logistic Biplots for Binary Data}
Let ${{\bf{X}}_{I \times J}}$  be a data matrix in which the rows correspond to I individuals and the columns to J binary characters. Let ${\pi_{ij}} = E({x_{ij}})$ the expected probability that the character $j$ be present at individual $i$, and ${x_{ij}}$ the observed probability, either 0 or 1, resulting in a binary data matrix. The S-dimensional logistic biplot  in the $logit $ scale is formulated as
\begin{equation}
logit({\pi_{ij}}) = \log ({{{\pi_{ij}}} \over {1 - {\pi_{ij}}}}) = {b_{j0}} + \sum\limits_{s = 1}^S {{b_{js}}{a_{is}}}  = {b_{j0}} + {{\bf{a'}}_i}{{\bf{b}}_j},
\label{binlogbip}
\end{equation}
where ${a_{is}}$  and ${b_{js}}$, $(i=1, \dots ,I; j=1, \dots ,J; s=1, ..., S)$, are the model parameters used as row and column markers respectively. The model is a generalized (bi)linear model having the $logit$ as a link function. In terms of probabilities rather than $logits$
\begin{equation}
{\pi_{ij}} = {{{e^{{b_{j0}} + \sum\nolimits_k {{b_{jk}}{a_{ik}}} }}} \over {1 + {e^{{b_{j0}} + \sum\nolimits_k {{b_{jk}}{a_{ik}}} }}}}
\label{binlogbip2}
\end{equation}
 In matrix form,  
  \begin{equation}
 logit({\bf{\Pi}}) = {{\bf{1}}_I}{{\bf{b'}}_0} + {\bf{AB'}},
 \label{binlogbip3}
 \end{equation}
where $\bf{\Pi}$ is the matrix of expected probabilities, ${{\bf{1}}_I}$ is a vector of ones and  ${{\bf{b}}_0}=(b_{j0})$  is the vector containing intercepts that have been added because it is not possible to center the data matrix in the same way as in linear biplots. The intercepts are the displacements of centroids in the same way as it is the first ordination axis in Correspondence Analysis. The model is a latent trait model for binary data, being the row coordinates the scores of individuals on the latent trait. Although the biplot in the logit scale may be useful, it would be more interpretable in a probability scale.
 
 The points predicting different probabilities are on parallel straight lines on the biplot; this means that predictions on the logistic biplot are made in the same way as on the linear biplots, i. e., projecting a row marker ${{\bf{a}}_i} = ({a_{i1}},{a_{i2}})$ onto a column marker ${{\bf{b}}_j} = ({b_{j1}},{b_{j2}})$.  (See \cite{Vicente-06}, \cite{Demeyetal}).

The model in (\ref{binlogbip}) is also a latent trait or item response theory model, in that ordination axes are considered as latent variables that explain the association between the observed variables. In this framework we suppose that individuals respond independently to variables, and that the variables are independent for given values of the latent traits.
With these assumptions the likelihood function is
 \begin{equation}
 {\rm{Prob}}(\left. {{x_{ij}}} \right|({{\bf{b}}_0},{\bf{A}},{\bf{B}})) = \prod\limits_{i = 1}^I {\prod\limits_{j = 1}^J {\pi _{ij}^{{x_{ij}}}{{(1 - {\pi _{ij}})}^{1 - {x_{ij}}}}} }
 \label{binlikelyhood}
 \end{equation}
Taking the logarithm of the likelihood function yields
 \begin{equation}
 L = {\rm{log}}\;{\rm{Prob}}(\left. {{x_{ij}}} \right|({{\bf{b}}_0},{\bf{A}},{\bf{B}})) = \sum\limits_{i = 1}^I {\sum\limits_{j = 1}^J {\left[ {{x_{ij}}\log ({\pi _{ij}}) + (1 - {x_{ij}})\log (1 - {\pi _{ij}})} \right]} }
  \label{binloglikelyhood}
  \end{equation}
  
For $\bf{A}$ fixed, (\ref{binloglikelyhood}) can be separated into $J$ parts, one for each variable,
   \begin{equation}
   L = \sum\limits_{J = 1}^J {{L_j}}  = \sum\limits_{J = 1}^J {\left( {\sum\limits_{i = 1}^I {\left[ {{x_{ij}}log({\pi _{ij}}) + (1 - {x_{ij}})log(1 - {\pi _{ij}})} \right]} } \right)} 
  \end{equation}
Maximizing each ${{L_j}}$ is equivalent to performing a standard logistic regression using the $j-th$ column of $\bf{X}$ as a response and the columns of $\bf{A}$ as regressors. 
In the same way the probability function can be separated into several parts, one for each row of the data matrix, $L = \sum\nolimits_{i = 1}^I {{L_i}} $.
 
Binary logistic biplots can be calculated using the package MULTBIPLOT \citep{MULTBIPLOT}. 

\section{Logistic Biplot for Nominal Data}

\subsection{Formulation}
Let ${{\bf{X}}_{I \times J}}$ be a data matrix containing the values of $J$ nominal variables, each with $K_j$ $(j=1, \ldots, J)$ categories, for $I$ individuals, and let ${{\bf{G}}_{I \times L}}$  be the corresponding indicator matrix with  $L = \sum\nolimits_j {{K_j}} $ columns.  The last (or the first) category of each variable will be used as a baseline. Let ${\pi _{ij(k)}}$ denote the expected probability that the category $k$ of variable $j$ be present at individual $i$. 
A multinomial logistic latent trait model with $S$ latent traits, states that the probabilities are obtained as
\begin{equation}
{\pi _{ij(k)}} = {{{e^{{b_{j(k)0}} + \sum\limits_{s = 1}^S {{b_{j(k)s}}{a_{is}}} }}} \over {\sum\limits_{l = 1}^{{K_j}} {{e^{{b_{j(l)0}} + \sum\limits_{s = 1}^S {{b_{j(l)s}}{a_{is}}} }}} }}, (k = 1, \ldots ,{K_j})
  \label{NominalProb}
\end{equation}
Using the last category as a baseline in order to make the model identifiable, the parameter for that category are restricted to be 0, i.e., ${b_{j({K_j})0}} = {b_{j({K_j})s}} = 0$, $(j = 1, \ldots ,J;\quad s = 1, \ldots ,S)$.The model can be rewritten as 
\begin{equation}
{\pi _{ij(k)}} = {{{e^{{b_{j(k)0}} + \sum\limits_{s = 1}^S {{b_{j(k)s}}{a_{is}}} }}} \over {1 + \sum\limits_{l = 1}^{{K_j} - 1} {{e^{{b_{j(l)0}} + \sum\limits_{s = 1}^S {{b_{j(l)s}}{a_{is}}} }}} }}, (k = 1, \ldots ,{K_j} - 1)
  \label{NominalProb2}
\end{equation}
With this restriction we assume that the log-odds of each response (relative to the last category) follows a linear model 
$$\log \left( {{{{\pi _{ij(k)}}} \over {{\pi _{ij({K_j})}}}}} \right) = {b_{j(k)0}} + \sum\limits_{s = 1}^S {{b_{j(k)s}}{a_{is}}}  = {b_{j(k)0}} + {{{\bf{a'}}}_i}{{\bf{b}}_{j(k)}},$$
where ${{a_{is}}}$  and ${b_{j(k)s}}\quad (i = 1, \ldots ,I;\quad j = 1, \ldots ,J;\quad k = 1, \ldots ,{K_j} - 1;\quad s = 1, \ldots ,S)$ are the model parameters. In matrix form,
\begin{equation}
{\bf{O}} = {{\bf{1}}_I}{{{\bf{b'}}}_0} + {\bf{AB'}},
  \label{NominalOdds}
\end{equation}
where ${{\bf{O}}_{I \times (L - J)}}$ is the matrix containing the expected log-odds, defines a biplot for the odds. Although the biplot for the odds may be useful, it would be more interpretable in terms of predicted probabilities and categories. This Biplot will be called ``Nominal Logistic Biplot'', and it is related to the latent nominal models in the same way as classical linear biplots are related to Factor or Principal Components Analysis or Binary Logistic Biplots are related to the Item Reponse Theory or Latent Trait Analysis for Binary data.

The points predicting different probabilities are no longer on parallel straight lines (see the figure 1 with the response surfaces); this means that predictions on the logistic biplot are not made in the same way as in the linear biplots, the surfaces define now prediction regions for each category as shown in the graph.

\subsection{Geometry}
\begin{figure}[!htb]
   \centering
   \subfloat[(a)]{\label{fig:a} \includegraphics[width=0.50\textwidth]{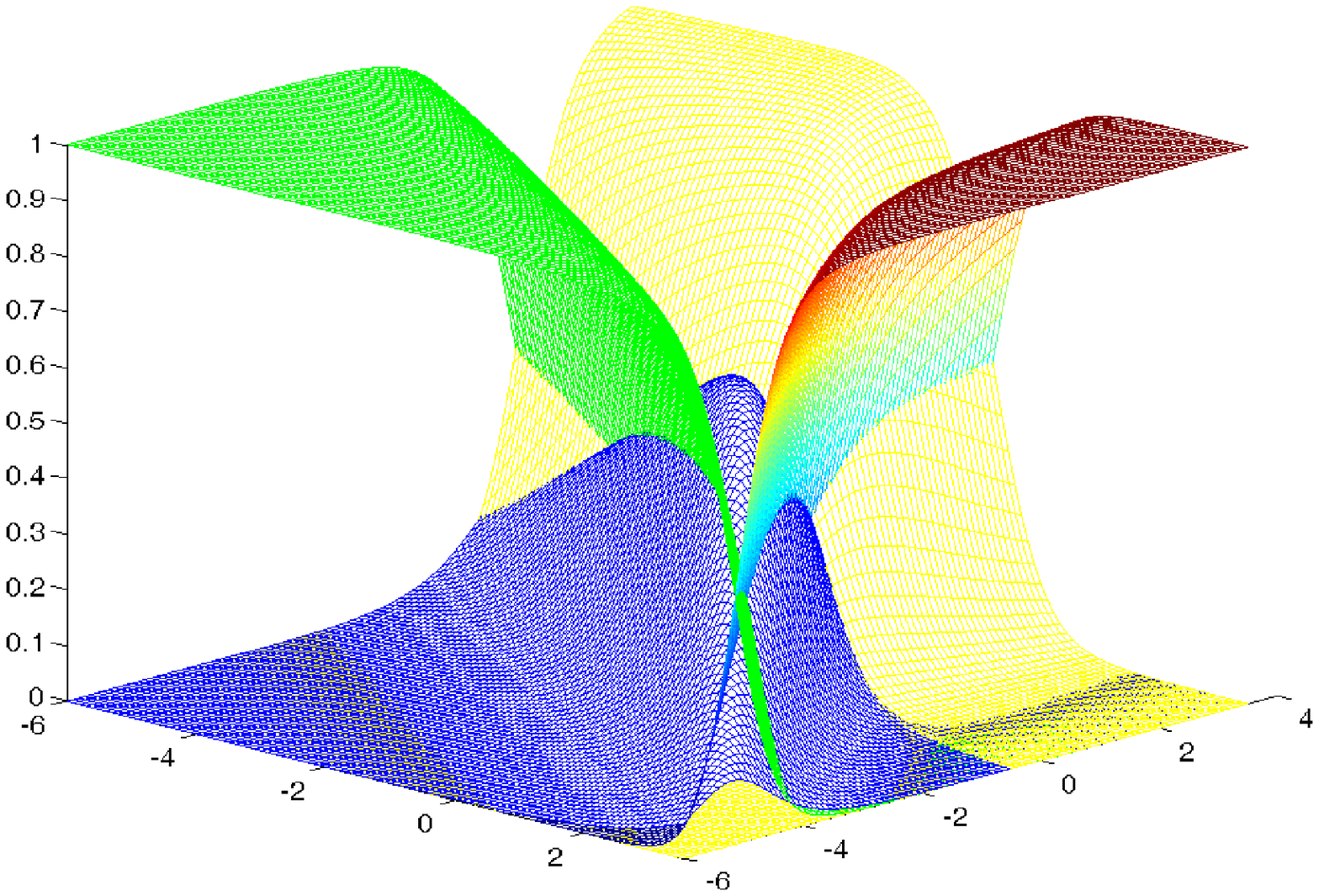}}
   \subfloat[(b)]{\label{fig:b} \includegraphics[width=0.50\textwidth]{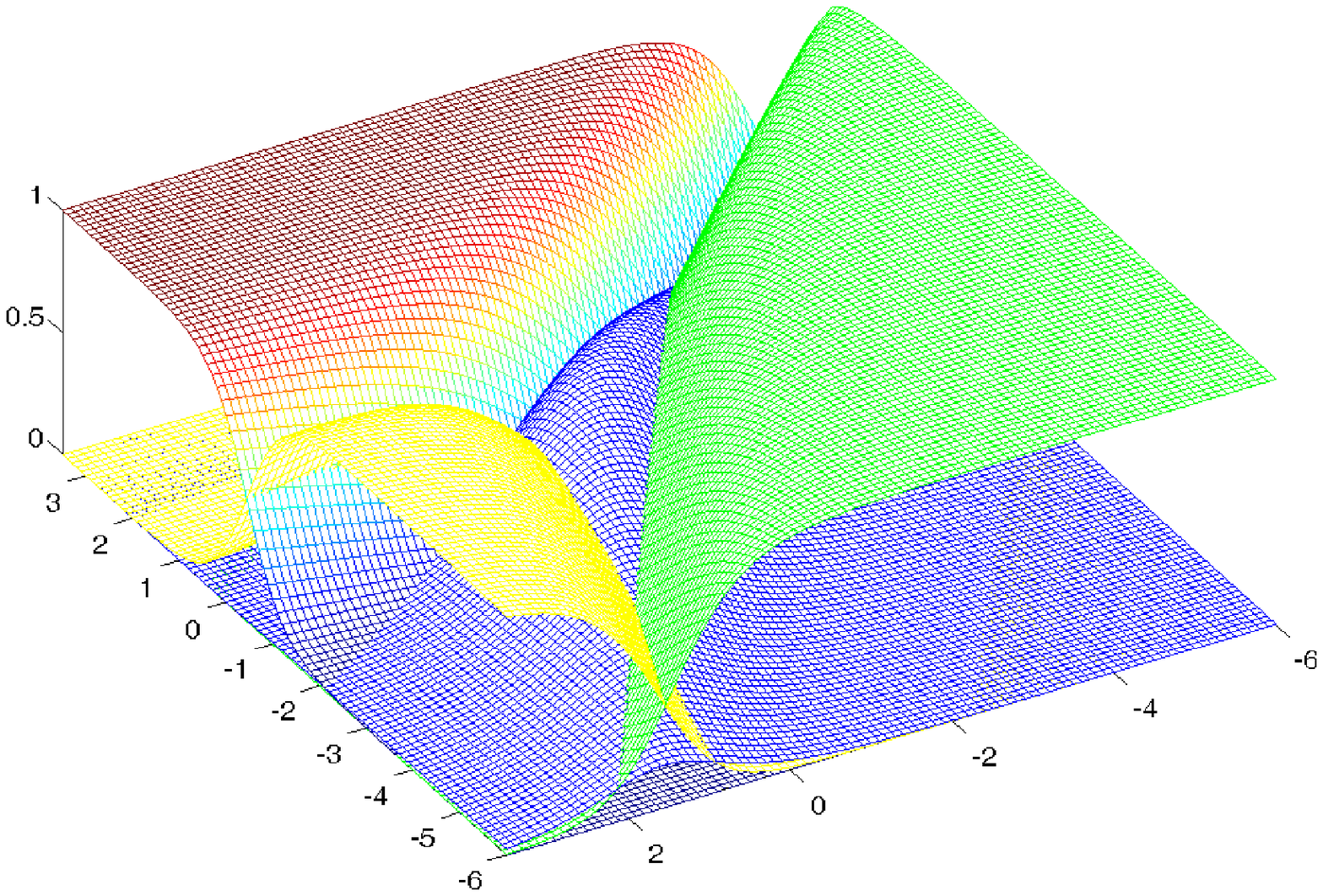}}\\[4pt]
   \caption{Response surfaces of the Nominal Logistic model (a,b) , with 4 categories and 2 explanatory variables.}
   \label{curvaslog}     %% Etiqueta para la figura entera
\end{figure}

Suppose we have a two-dimensional representation in which the row coordinates are defined by the first two columns of ${\bf{A}}$ in (\ref{NominalOdds}), let's call $\mathcal{L}$ the space generated by those columns. Equations (\ref{NominalProb}), (\ref{NominalProb2}) and (\ref{NominalOdds}) define a set of probability response surfaces (one for each category and each variable) (figure \ref{curvaslog}) that are no longer sigmoid as in the binary case (\cite{Vicente-06}). This means that the level curves are no longer straight lines and then, prediction of probabilities is not made by projection as in the usual linear biplots. Figure~\ref{fig2:b} shows the level curves for probability 0.5 and a hypothetical variable with four categories. We will show that in this case the predicted probabilities, for each variable, define a set of convex polygons that can be interpreted as ``prediction'' regions in the same way as in \cite{Gower-96}. For each variable there are as many regions as categories and each one is formed by the set points in with the expected probability for a category is higher than the probability for the rest of  categories. Let ${\mathcal{R}_k}$ denote the region for category $j$, then it can be defined as
\begin{equation*}
{\mathcal{R}_k} = \left\{ {{{\bf{a}}_h} = ({a_{h1}},{a_{h2}}) \in \mathcal{L}/{\pi _{hj(k)}} \ge {\pi _{hj(m)}},\forall m \ne k;k,m = (1, \ldots ,{K_j})} \right\}
\end{equation*}
The prediction regions for a hypothetical variable with four categories are shown in figure~\ref{fig2:c}. It is immediate to see that the prediction regions are closely related to the level curves. 

It has to be noted that there are some cases in which some of the categories are never predicted, those will be termed \textbf{hidden categories} and should be taken into account to construct the final representation.

\subsection{Obtaining ``prediction regions''}
In the following paragraphs we will describe a procedure to obtain the prediction regions using methods taken from the Computational Geometry. The set of convex polygons predicting each category form a tessellation of the plane. Each cell of the tessellation is delimited by a set of straight lines that correspond to points that have equal probabilities for two of the categories of the variable (the edges). We consider each variable $j$ ($j = 1, \ldots ,J$) separately. Each pair of response surfaces defined by (\ref{NominalProb}) intersect in a straight line that, projected onto the space of predictors, is the set of points in which the probability of both categories is the same. Those lines are the candidates to be the edges of the convex polygons defining the prediction regions. That is, we search for the set of points $\mathcal{E}_{kl}$ in $\mathcal{L}$ such that the pair of categories $k$ and $l$ ($k,l = 1, \ldots ,{K_j}$), have the same expected probability $\pi_{ij(k)}=\pi_{ij(l)}$ i. e., $\mathcal{E}_{kl}$ is the set of points verifying:

\begin{equation}
{{{e^{{b_{j(k)0}} + \sum\limits_{s = 1}^2 {{b_{j(k)s}}{a_{s}}} }}} \over {\sum\limits_{m = 1}^{{K_j}} {{e^{{b_{j(m)0}} + \sum\limits_{s = 1}^2 {{b_{j(m)s}}{a_{s}}} }}} }} = {{{e^{{b_{j(l)0}} + \sum\limits_{s = 1}^2 {{b_{j(l)s}}{a_{s}}} }}} \over {\sum\limits_{m = 1}^{{K_j}} {{e^{{b_{j(m)0}} + \sum\limits_{s = 1}^2 {{b_{j(m)s}}{a_{s}}} }}} }}
  \label{LinePair}
\end{equation}

Then
\begin{equation*}
{b_{j(k)0}} + \sum\limits_{s = 1}^2 {{b_{j(k)s}}{a_{s}}} = {b_{j(l)0}} + \sum\limits_{s = 1}^2 {{b_{j(l)s}}{a_{s}}}
\end{equation*}
or
\begin{equation*}
(b_{j(k)1}-b_{j(l)1}) a_{1} + (b_{j(k)2}-b_{j(l)2}) a_{2}  = (b_{j(l)0}-b_{j(k)0})
\end{equation*}

\begin{figure}[!htb]
   \subfloat[(a)]{\label{fig2:a} \includegraphics[width=0.40\textwidth]{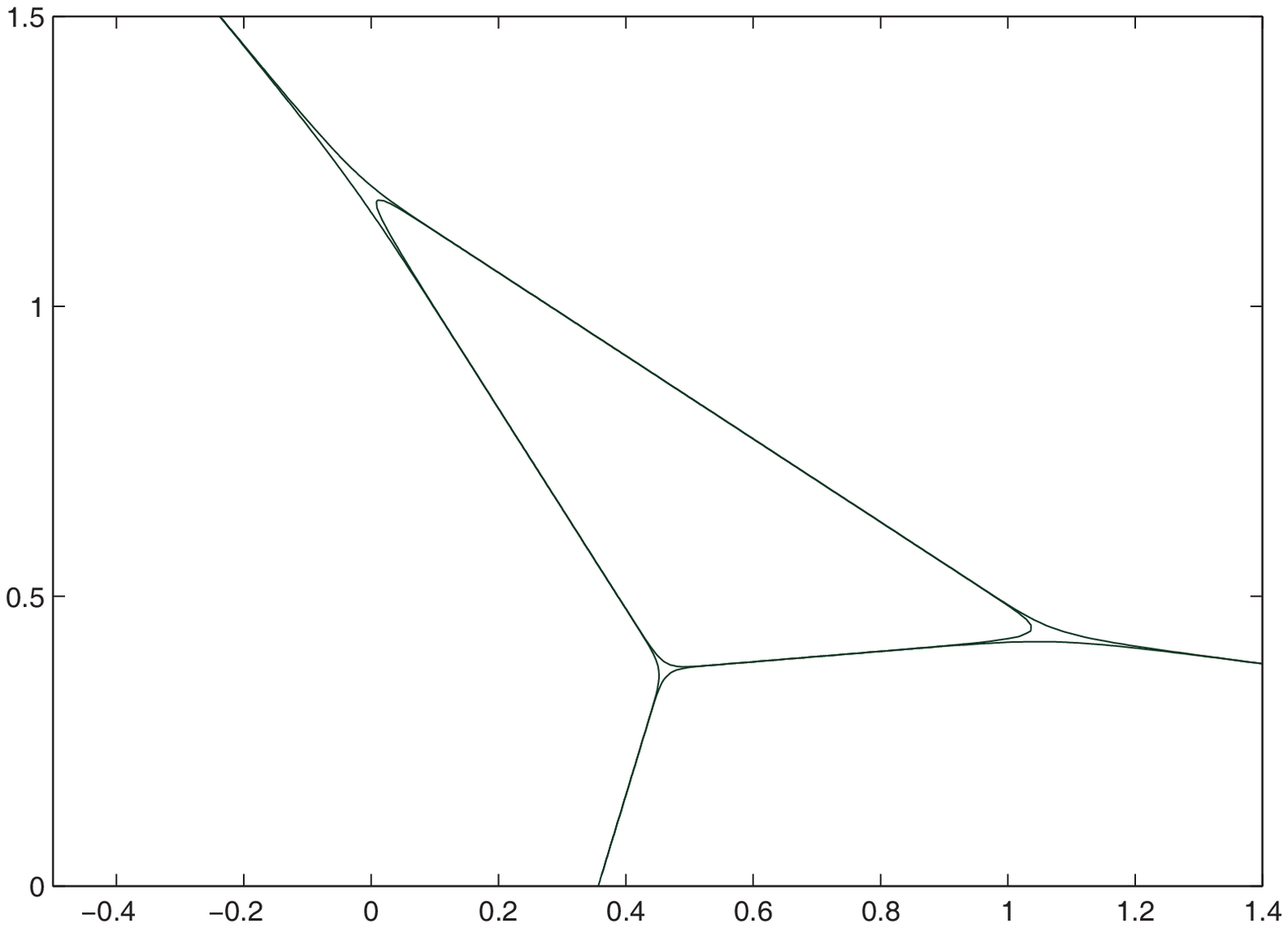}}
   \subfloat[(b)]{\label{fig2:b} \includegraphics[width=0.40\textwidth]{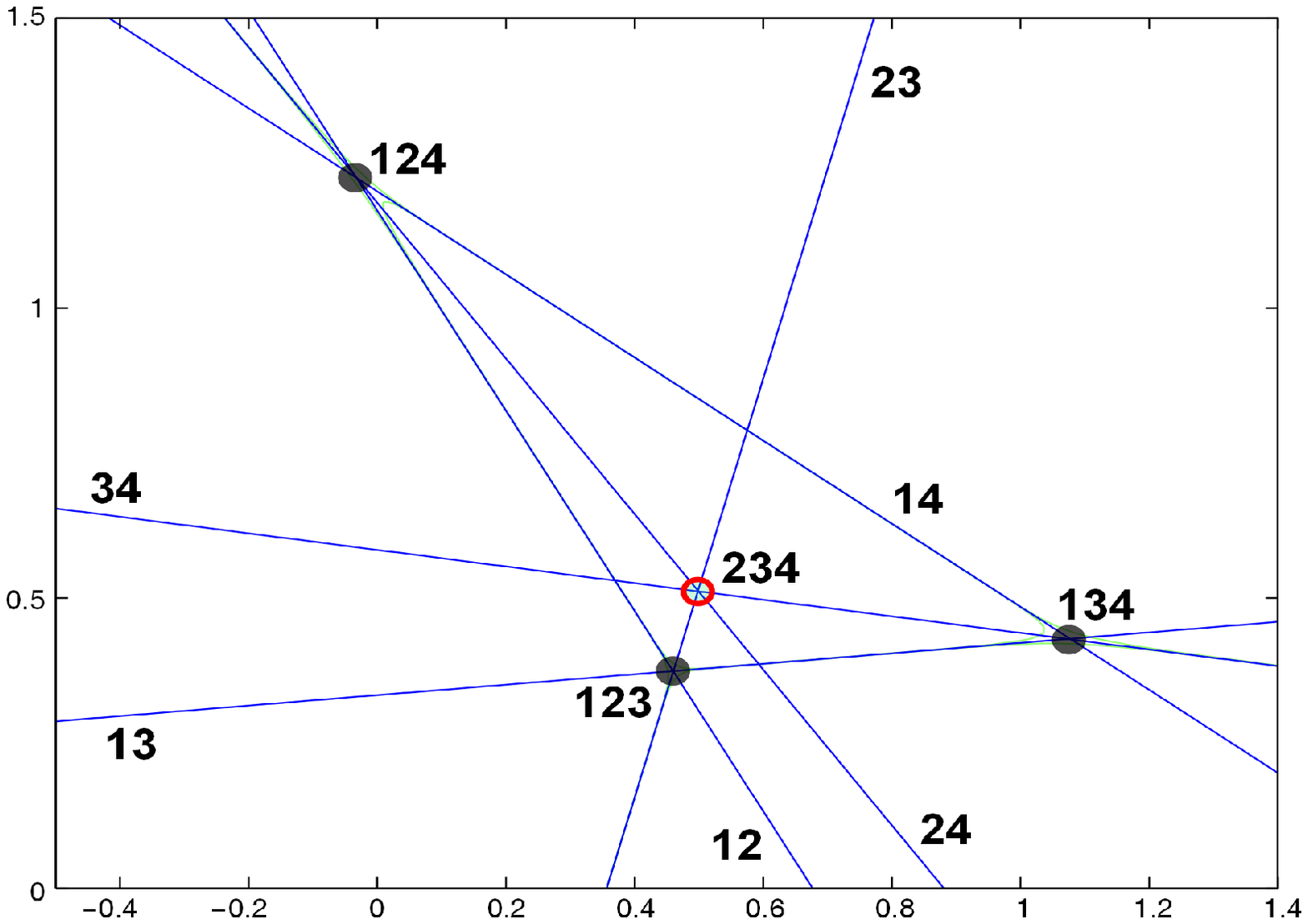}}
   \\[4pt]
   \centering   
   \subfloat[(c)]{\label{fig2:c}
   \includegraphics[width=0.40\textwidth]{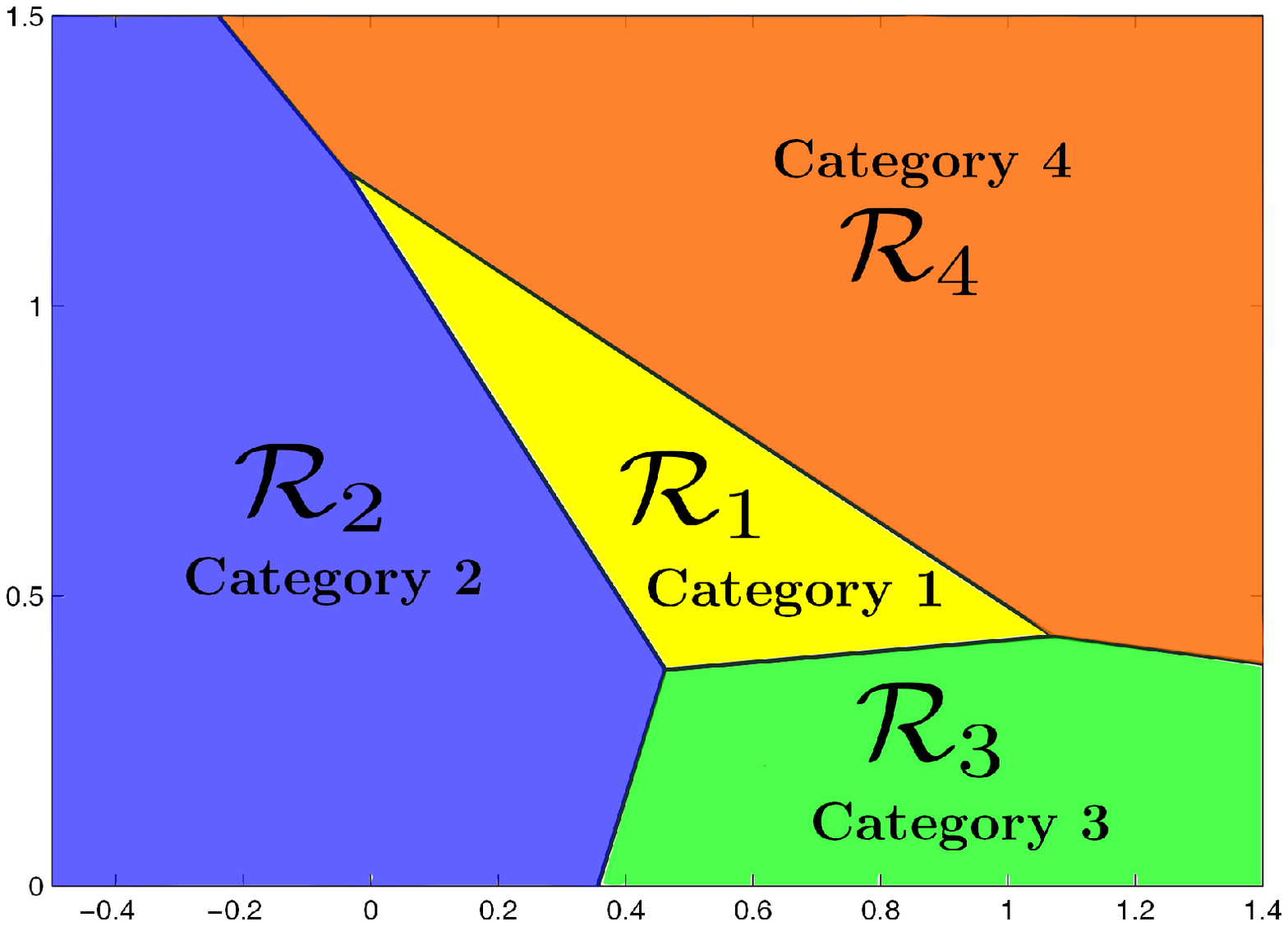}}\\[4pt]
   \caption{Geometry for a two-dimensional solution and a hypothetical variable with four categories:  Level curves of the response surfaces for  $p=0.5$ (a), lines of equal probability for each pair of categories and their intersection points (candidates for edges and vertices of the tessellation) (b) and tessellation of the plane defined by the prediction regions (c).}
   \label{teselarectas}     %% Etiqueta para la figura entera
\end{figure}

The above equation can be written as:
\begin{equation*}
{a_2} = {{({b_{j(l)0}} - {b_{j(k)0}})} \over {({b_{j(k)2}} - {b_{j(l)2}})}} - {{({b_{j(k)1}} - {b_{j(l)1}})} \over {({b_{j(k)2}} - {b_{j(l)2}})}}{a_1},
\end{equation*}
where ${a_1}$ and ${a_2}$ are generic coordinates on the dimensions of $\mathcal{L}$. 
Each variable $j$ has  ${K_j \choose 2}$ of such lines as shown in figure  \ref{fig2:b} for a hypothetical example with four categories.

\begin{figure}[!htb]
 \centering
 \includegraphics[width=0.5\textwidth]{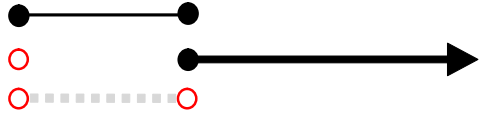}
 \caption{Definition of join for constructing the tessellation. Black $\bullet$: real point;  Red $\odot$: virtual point}
 \label{realvirtual}     %% Etiqueta para la figuraentera
\end{figure}

Except for degenerate cases, any two lines with one index in common, $\mathcal{E}_{kl}$ and $\mathcal{E}_{km}$, intersect in a point $\mathcal{P}_{klm}$. The ${K_j \choose 3}$ of such points are the candidates to be the vertices of the tessellation.  Point $\mathcal{P}_{klm}$ is a vertex of the tessellation if there is not a $t \notin \left\{ {k,l,m} \right\}$ such that ${\pi _{({\mathcal{P}_{klm}})t}} > {\pi _{({\mathcal{P}_{klm}})r}}$ for $r \in \left\{ {k,l,m} \right\}$, where ${\pi _{({\mathcal{P}_{klm}})t}}$ is the expected probability of category $t$ at point $\mathcal{P}_{klm}$, i.e., the expected probability for one of the categories involved is the highest. If a point is a vertex of the tessellation is termed \textbf{real point}, otherwise is a \textbf{virtual point}. Degenerate cases may have parallel lines but this is extremely unlikely to occur. 
The prediction regions ${\mathcal{R}_k}$ are delimited by all the lines  $\mathcal{E}_{kl}$ with index $k$ and its vertices are all the points $\mathcal{P}_{klm}$ with the index $k$. A category is hidden when its index is not present in any of the real points. The region of the hidden category is omitted in the representation.
We now define the meaning of \textbf{join} two points $\mathcal{P}_{klm}$ and $\mathcal{P}_{kln}$ as follows (see figure  \ref{realvirtual}):
\begin{enumerate}
\item Two real points should be joined, if they have two indices in common, following the line $\mathcal{E}_{kl}$.
\item Two virtual points are never joined.
\item A virtual point and a real point are joined along the line $\mathcal{E}_{kl}$, starting from the real point and away from the virtual point.
\end{enumerate}

Now it is easy to adapt the algorithm described in \cite{Gower-96} to construct the tessellation generated by the probability responses:
\begin{enumerate}
\item Compute the coordinates of all ${K_j \choose 3}$ points $\mathcal{P}_{klm}$.
\item Decide if the point is real or virtual.
\item Join all pairs of points sharing two suffices, interpreting ``join'' as described before.
\end{enumerate}

The procedure is different from that in  \cite{Gower-96} in two aspects: They start from a set of points $\mathcal{C}_{k}, k=(1, \ldots, K_j)$ that they call ``category points'' arising from a Multiple Correspondence Analysis with some modifications, and then construct the tessellation from those points using distances; we don't have the category points and use probabilities rather than distances. The tessellation based on distances is called a Voronoi Diagram and is quite a popular tool in a discipline called ``Computational Geometry''; in this diagram the space is divided into a set of polygons or regions ${\mathcal{R}_k}$ in such a way that points in the region are closest to  $\mathcal{C}_{k}$ than to any other point. The main advantage of doing so is that it provides a simple interpretation of the representation of row and column markers of the data matrix, the predicted category for each point is the corresponding to its closest ``category point''. Representing points rather than ``regions'' produces a much cleaner  and easier way to interpret the graph.
We have the regions but not the points and, although from a formal point of view our problem is solved, and we have a simultaneous representation of individuals and variables, it would be more convenient to have also a set of ``category points'' to interpret the biplot in terms of distances. Let's call this set of points ${\mathcal{C}_{j(k)}},j = (1, \ldots ,J),k = (1, \ldots ,{K_j})$. This would be a fundamental contribution of our research becase the interpretation of distances among row and column points is simple and it is not an intrinsic property of most multivariate techniques as MCA, except Unfolding that is designed for a completely different purpose. 
Three problems arise:
\begin{enumerate}
\item Is our tessellation a Voronoi diagram?.
\item If not, is there any way to approximate it by its closest Voronoi tessellation?.
\item Given a Voronoi tessellation, is it possible to obtain a set of generators for it?.
\end{enumerate}
In the next section we describe a procedure to obtain the generators given a tessellation.

\subsection{Obtaining generators of the tessellation}
\begin{figure}[!htb]
 \centering
 \includegraphics[width=0.7\textwidth]{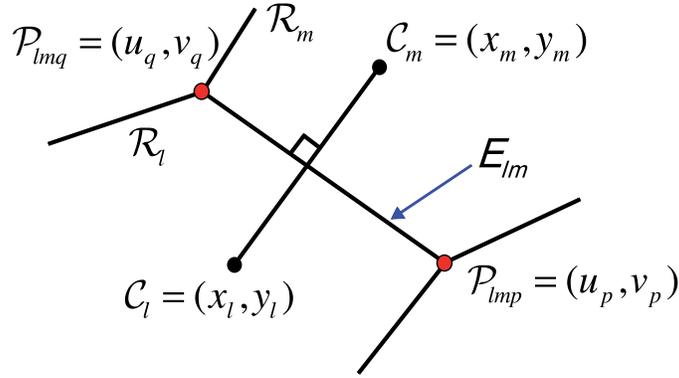}
 \caption{Centers ${{\cal C}_l} = ({x_l},{y_l})$ and ${{\cal C}_m} = ({x_m},{y_m})$ are equidistant from the edge they share $E_{lm}$ (\ref{arista}) and both lie on the line perpendicular to $E_{lm}$ (\ref{perpen})}
 \label{centrosvor}     %% Etiqueta para la figuraentera
\end{figure}

The problem of testing if any convex tessellation consists of Voronoi polygons and if so, obtain a set of centers or generators of the Voronoi diagram, has been studied for example by \cite{Hartvigsen-92} and \cite{Evans-87}. The first paper establishes a set of equations of slope and distance that a tesselation must hold to be Voronoi in such a way that solving a linear system it is possible to obtain the set of centers(figure \ref{centrosvor}). Let's see it in more detail.

First consider the following result (we will omit the index $j$ of the variable for simplicity): A tessellation of  $K$ polygons or convex regions  $\mathcal{R}_{k}, k=1,\dots,K$ is a Voronoi diagram with centres $\mathcal{C}_k = ({x_{k}},{y_{k}}),k = (1, \ldots ,{K})$ iff  $\mathcal{R}_{k}=\{(x,y):(x-{x_{k}})^2 + (y-{y_{k}})^2 \leq (x-{x_{l}})^2 + (y-{y_{l}})^2 ,\forall l \neq k\}$, i.e., each polygon of the tessellation is the set of points that are nearer to its center than to any center of other polygon.

 If we consider two adjacent polygons, $\mathcal{R}_{l}$ y $\mathcal{R}_{m}$, whose common edge is  $E_{lm}$ with equation $y=s_{i}x + b_{i}$, and contain the vertices $(u_{p},v_{p})$ and $(u_{q},v_{q})$, let ${{\cal C}_l} = ({x_l},{y_l})$ and ${{\cal C}_m} = ({x_m},{y_m})$ the Voronoi centers of the regions (our ``category points'').  The equations of slope and distance are:
\begin{equation}
\frac{({y}_{l}-{y}_{m})}{({x}_{l}-{x}_{m})}=\frac{-1}{s_i}
\label{arista}
\end{equation}
\begin{equation}
-s_{i}{x_l} + {y_l} - b_i = -s_{i}{x}_{m} + {y}_{m} - b_i 
\label{perpen}
\end{equation}
where $s_{i}=\frac{(v_{p}-v_{q})}{(u_{p}-u_{q})}$ and $b_{i}=s_{i}u_{p} -v_{p}$.

Those equations with, for example $k$ edges and  $n$ polygons form a linear system with $2k$ equations and  $2n$ unknowns, that can be solved by least squares. In matrix form the system is

\begin{equation*}
\begin{split}
&Bx=0\\
&Ax=b,
\end{split}
\end{equation*}
with $x=[x_{1},y_{1},\dots,x_{n},y_{n}]'$, $b=-2[b_{1},\dots,b_{k}]'$. Matrices $A$ y $B$ are sparse but that is not a problem because the number of categories is usually small. Calculations to obtain a solution are based on three algorithms that can produce different centers in the case that the polygons of the tessellation are not Voronoi. The three methods are:

\textbf{Algorithm 1}: Minimize the conditions of distance and slope, that is, search for  $Min\left \| \begin{bmatrix}{A}\\{B}\end{bmatrix} x - \begin{bmatrix}{b}\\{0}\end{bmatrix}  \right \|^2$ , with $\left \| . \right \|^2$  the euclidean norm.

\textbf{Algorithm 2}: Minimize $\left \| Bx  \right \|^2$ , subject to $Ax=b$.

\textbf{Algorithm 3}: Minimize $\left \| Ax - b  \right \|^2$ , subject to $Bx=0$.

\begin{figure}[!htb]
   \centering
   \subfloat[(a)]{\label{perfil:a} \includegraphics[width=0.45\textwidth]{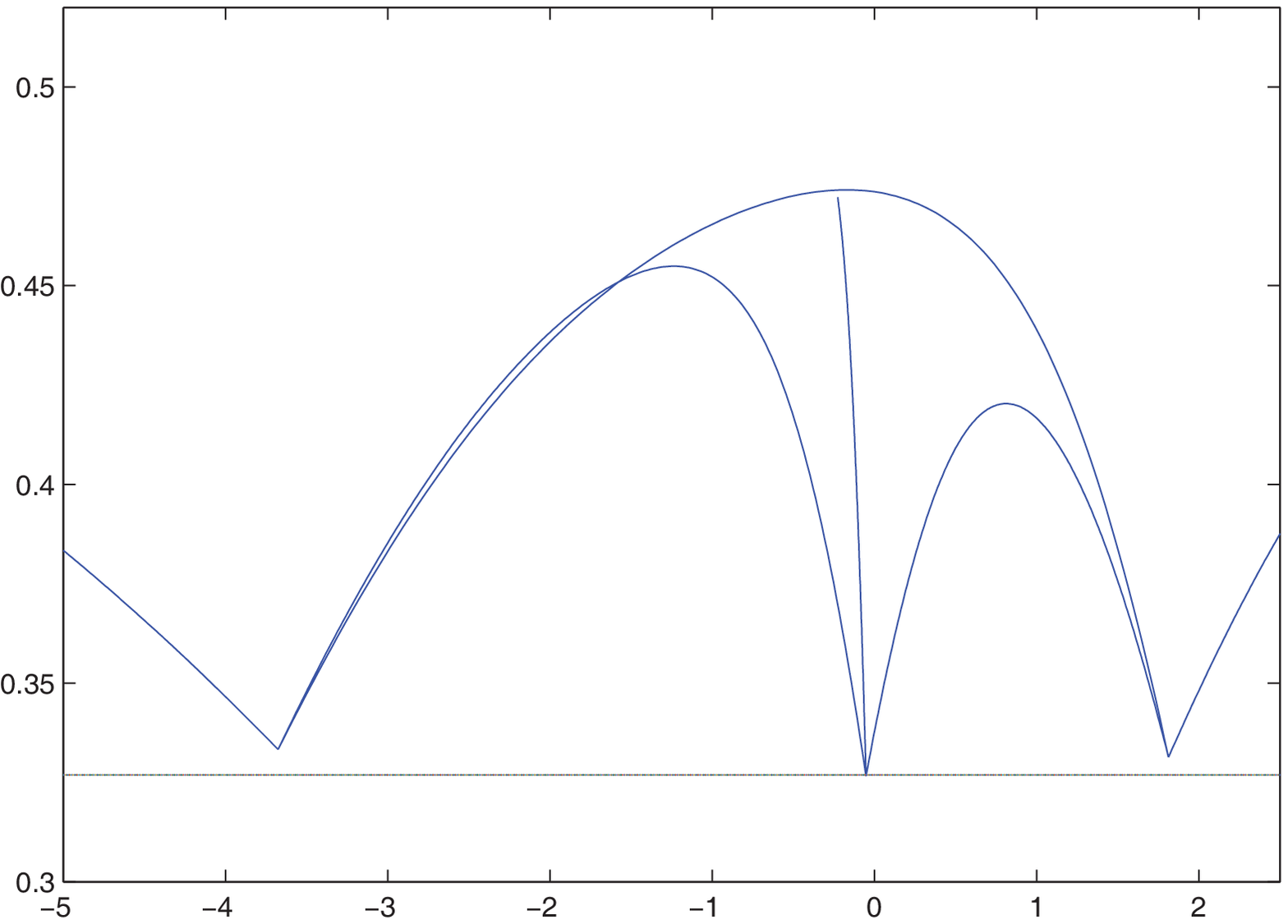}}
   \subfloat[(b)]{\label{planta:b} \includegraphics[width=0.45\textwidth]{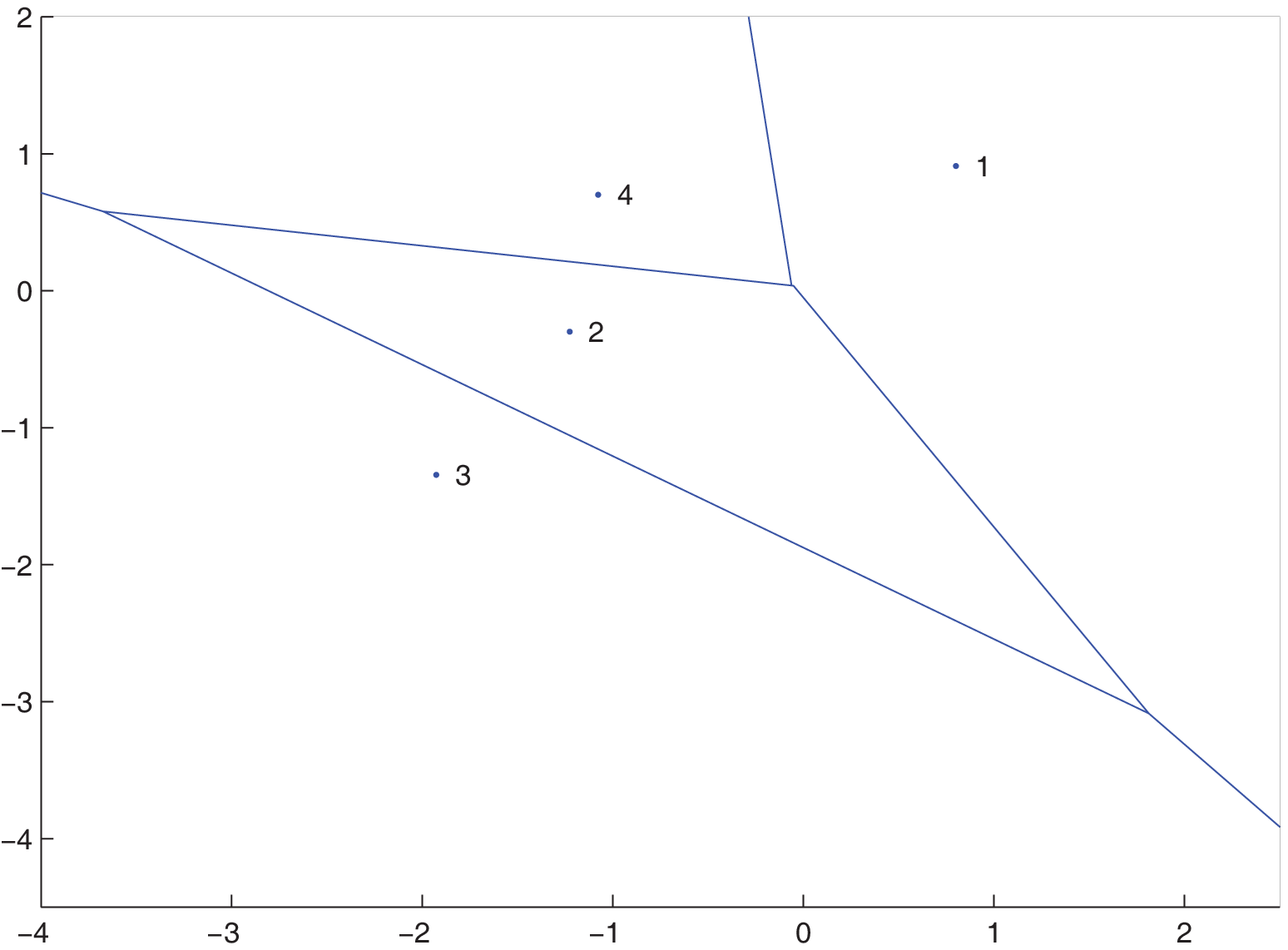}}\\[4pt]
   \caption{Frontal view of the intersections among the 4 response curves obtained from the Nominal Logistic Regression (a) and the generated tessellation with the result of the algorithm for inversion (b) with the category points.}
   \label{perfilplanta}     %% Etiqueta para la figura entera
\end{figure}

In practice, the main problem with the linear systems is the instability of the algorithms due to the ill conditioning of the matrices. \cite{Schoenberg-03} treats the problem and propose some alternatives to improve the stability of the final solution.  \cite{Evans-87} also proposes a measure of the goodness of fit, i. e., a measure of how near is the tessellation from a true Voronoi diagram. 
For the hypothetical example in figure  \ref{curvaslog} we show the result of inverting a tessellation obtained from the logistic response in figure \ref{perfilplanta}, for this case the tessellation is very close to a Voronoi diagram. 

\subsection[GNLB]{Paramater estimation}
Although the nominal case doesn't share the geometrical properties with the binary case, the alternated algorithm described in \cite{Vicente-06}, can be easily extended replacing the binary logistic regressions by multinomial logistic regressions. The problem with this approach is that the parameters for the individuals can not be estimated when the individual has 0 or 1 in all the variables for the binary case, or all the responses are at the baseline category for the nominal case. In this paper we use a procedure that is similar to the alternated regressions method, except that the interpolation step is ``eliminated'' by considering the row parameters as incidental.  The technique assumes that the scores for individuals are random effects sampled from some larger distribution. The estimation procedure is an EM-algorithm that uses the Gauss-Hermite quadrature to approximate the integrals, considering the individual scores as missing data.  More details of similar procedures can be found in \cite{Bock81} or \cite{Chalmers2012}.

The likelihood function is
$$M\left( {\left. {\bf{G}} \right|{{\bf{b}}_0},{\bf{A}},{\bf{B}}} \right) = \prod\limits_{i = 1}^I {\prod\limits_{j = 1}^J {\prod\limits_{k = 1}^{{K_j}} {\pi _{ij(k)}^{{g_{ij(k)}}}} } }, $$
where ${{g_{ij(k)}}}=1$  if individual $i$ chooses category $k$ of item $j$ and ${{g_{ij(k)}}}=0$ otherwise. 
The log-likelihood is

\begin{equation}
L\left( {\left. {\bf{G}} \right|{{\bf{b}}_0},{\bf{A}},{\bf{B}}} \right) = \sum\limits_{i = 1}^I {\sum\limits_{j = 1}^J {\sum\limits_{k = 1}^{{K_j}} {{g_{ij(k)}}} } \log \left( {{\pi _{ij(k)}}} \right)}
\label{LogLikNom}
\end{equation}

If the parameters $\bf{A}$ for  individuals where known, the log-likelihood could be separated into $J$ parts, one for each variable

\begin{equation}
L\left( {\left. {\bf{G}} \right|{{\bf{b}}_0},{\bf{B}}} \right) = \sum\limits_{j = 1}^J {{L_j}(\left. {\bf{G}} \right|{{\bf{b}}_{j0}},{{\bf{B}}_j})}  = \sum\limits_{j = 1}^J {\left( {\sum\limits_{i = 1}^I {\sum\limits_{k = 1}^{{K_j}} {{g_{ij(k)}}} } \log \left( {{\pi _{ij(k)}}} \right)} \right)},
\label{LogLikNom2}
\end{equation}
where ${{{\bf{b}}_{j0}}}$ and ${{{\bf{B}}_j}}$ are the submatrices of parameters for the $j$th variable. Maximizing the log-likelihood is equivalent to maximizing each part, i.e., obtaining the parameters for each variable separately. Maximizing each $L_j$ is equivalent to performing a multinomial logistic regression using the $j$th column of $\bf{X}$ as response and the columns of $\bf{A}$ as predictors. We do not describe logistic regression here because it is as a very well known procedure. It is also well-known that when the individuals for different categories are separated (or quasi-separated) on the space spanned by the explanatory variables, the maximum likelihood estimators does not exist (or are unstable). Because we are seen the biplot as a procedure to classify the set of individuals and searching for the variables responsible for it, accounting for as much of the information as possible, it is probable that, for some variables, the individual are separated and then the procedure does not work just because the solution is good.   The problem of the existence of the estimators in logistic regression can be seen in \cite{Albert1984}, a solution for the binary case, based on the Firth's method \citep{Firth1993} is proposed by \cite{Heinze2002}. The extension to nominal logistic model was made by \cite{Bull2002}. All the procedures were initially developed to remove the bias but work well to avoid the problem of separation. Here we have chosen a simpler solution based on ridge estimators for logistic regression \citep{Cessie1992}.  

Rather than maximizing ${{L_j}(\left. {\bf{G}} \right|{{\bf{b}}_{j0}},{{\bf{B}}_j})}$ we maximize
\begin{equation}
{{L_j}(\left. {\bf{G}} \right|{{\bf{b}}_{j0}},{{\bf{B}}_j})} - \lambda \left( {\left\| {{{\bf{b}}_{j0}}} \right\| + \left\| {{{\bf{B}}_j}} \right\|} \right)
\label{Penalized}
\end{equation}

We don't describe here the procedure in great detail because that is also a standard procedure. Changing the values of $\lambda$ we obtain slightly different solutions not affected by the separation problem.

In the same way, if  parameters for  variables were known, the log-likelihood could be separated into $I$ parts, one for each individual.
$$L\left( {\left. {\bf{G}} \right|{\bf{A}}} \right) = \sum\limits_{i = 1}^I {{L_i}(\left. {\bf{G}} \right|{{\bf{a}}_i})}  = \sum\limits_{i = 1}^I {\left( {\sum\limits_{j = 1}^J {\sum\limits_{k = 1}^{{K_j}} {{g_{ij(k)}}\log } } \left( {{\pi _{ij(k)}}} \right)} \right)} $$
To maximize each part we could use Newton-Raphson with  a penalization as before. Rather than that we will use expected a posteriori estimators for the individual markers. 
For each individual (or response pattern) ${{\bf{g}}_i}$, the likelihood is
$${M_\ell }\left( {\left. {{{\bf{g}}_i}} \right|{{\bf{b}}_0},{{\bf{a}}_i},{\bf{B}}} \right) = \prod\limits_{j = 1}^J {\prod\limits_{k = 1}^{{K_j}} {\pi _{ij(k)}^{{g_{ij(k)}}}} } $$
Assuming a distributional form $g({\bf{a}})$ (multivariate normal, for example) the marginal distribution becomes
$${P_l}\left( {\left. {{{\bf{b}}_0},{\bf{B}}} \right|{\bf{g}}} \right) = \int_{ - \infty }^\infty   \ldots  \int_{ - \infty }^\infty  {{M_\ell }\left( {\left. {{{\bf{g}}_i}} \right|{{\bf{b}}_0},{{\bf{a}}_i},{\bf{B}}} \right)} g({\bf{a}})d{\bf{a}},$$
the observed likelihood is
$$M\left( {\left. {{{\bf{b}}_0},{\bf{B}}} \right|{\bf{G}}} \right) = \prod\limits_{i = 1}^I {\left[ {\int_{ - \infty }^\infty   \ldots  \int_{ - \infty }^\infty  {{M_\ell }\left( {\left. {{{\bf{g}}_i}} \right|{{\bf{b}}_0},{{\bf{a}}_i},{\bf{B}}} \right)} g({\bf{a}})d{\bf{a}}} \right]} $$
We approximate the integral by $S$-dimensional Gauss-Hermite quadrature
$${{\tilde P}_l} = \sum\limits_{qS = 1}^Q { \ldots \sum\limits_{q1 = 1}^Q {{M_\ell }\left( {\left. {{{\bf{g}}_\ell }} \right|{{\bf{b}}_0},{\bf{Y}},{\bf{B}}} \right)g({y_{q1}})}  \ldots g({y_{qS}})} $$
The multivariate $S$-dimensional quadrature, \textbf{Y}, has been obtained as the product of $S$ unidimensional quadratures $({y_1}, \ldots ,{y_Q})$ with $Q$ nodes each. Then the marginal expected a posteriori score for individual $i$ at dimension $s$, ${a_{is}}$, is

$$E({a_s}/{{\bf{g}}_\ell }) = {{\sum\nolimits_{q = 1}^Q {{y_q}{M_\ell }\left( {\left. {{{\bf{g}}_\ell }} \right|{{\bf{b}}_0},{\bf{Y}},{\bf{B}}} \right)g({y_q})} } \over {{{\tilde P}_l}}}$$

\section{A real case}\label{example}

Table~\ref{data}, taken from \cite{Gower-96}, shows the observations of four variables observed on twenty farms from the Dutch island of Terschelling. This table is reported in \cite{Jongman-87} and it is part of a much larger survey. It is concerned with environmental factors and different forms of farm management. We have chosen this data because it has been previously analysed in literature and can serve as a comparison with the methods proposed here.

\begin{table}[!htb]

\centering\small \caption{Data on four variables observed at 20 farms on the island of Terschelling}\label{datafarm}
\renewcommand{\arraystretch}{0.9}
\begin{tabular}{ccccc}\hline
\parbox[c][0.4\height]{10mm}{\centering\smallskip \textbf\tiny{Farm\\number}\smallskip} &%
\parbox[c][0.5\height]{10mm}{\centering\smallskip \textbf\tiny{Moisture\\class}\smallskip} &%
\parbox[c][1\height]{15mm}{\centering\smallskip \textbf\tiny{Grassland\\management\\type}\smallskip} &%
\parbox[c][0.5\height]{15mm}{\centering\smallskip \textbf\tiny{Grassland\\use}\smallskip} &%
\parbox[c][0.5\height]{10mm}{\centering\smallskip \textbf\tiny{Manure\\class}\smallskip} \\\hline
1 & 1 &	SF &	2 &	4\\
2 & 1 &	BF &	2 &	2\\
3 & 2 &	SF &	2 &	4\\
4 & 2 &	SF &	2 &	4\\
5 & 1 &	HF &	1 &	2\\
6 & 1 &	HF &	2 &	2\\
7 & 1 &	HF &	3 &	3\\
8 & 5 &	HF &	3 &	3\\
9 & 4 &	HF &	1 &	1\\
10 & 2 & BF &	1 &	1\\
11 & 1 & BF &	3 &	1\\
12 & 4 & SF &	2 &	2\\
13 & 5 & SF &	2 &	3\\
14 & 5 & NM &	3 &	0\\
15 & 5 & NM &	2 &	0\\
16 & 5 & SF &	3 &	3\\
17 & 2 & NM &	1 &	0\\
18 & 1 & NM &	1 &	0\\
19 & 5 & NM &	1 &	0\\
20 & 5 & NM &	1 &	0\\
\hline
\end{tabular}
\label{data}
\end{table}

The variables are: 

\begin{itemize} 
\item Moisture class, with 5 levels, although level 3 does not occur in the data. Levels are labelled M1, M2, M4 and M5.
\item Grassland management type, with 4 levels (standard farming (SF), biological farming (BF), hobby farming (HF) and nature conservation management(NM))
\item Grassland use, with three levels: (production(U1), intermediate(U2) and grazing(U3))
\item Manure class, with 5 levels labelled C0, C1, C2, C3 and C4. The variable is probably ordinal because the levels assume an increasing level of manure but will be treated as categorical here.
\end{itemize}

\begin{figure}[!htb]
   \centering
   \subfloat[(a) Moisture]{\label{fig:a} \includegraphics[width=0.50\textwidth]{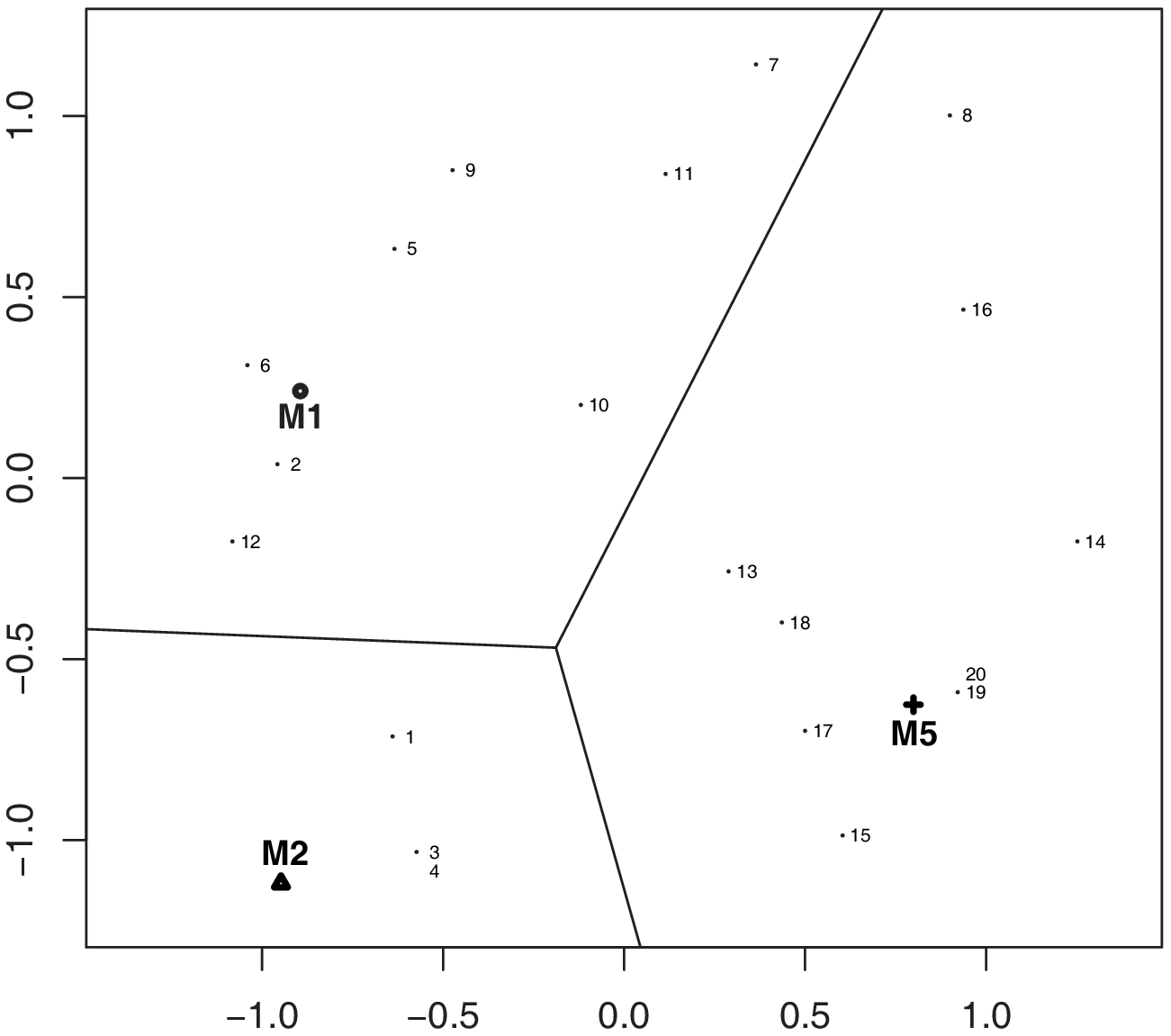}}
   \subfloat[(b) Management]{\label{fig:b} \includegraphics[width=0.50\textwidth]{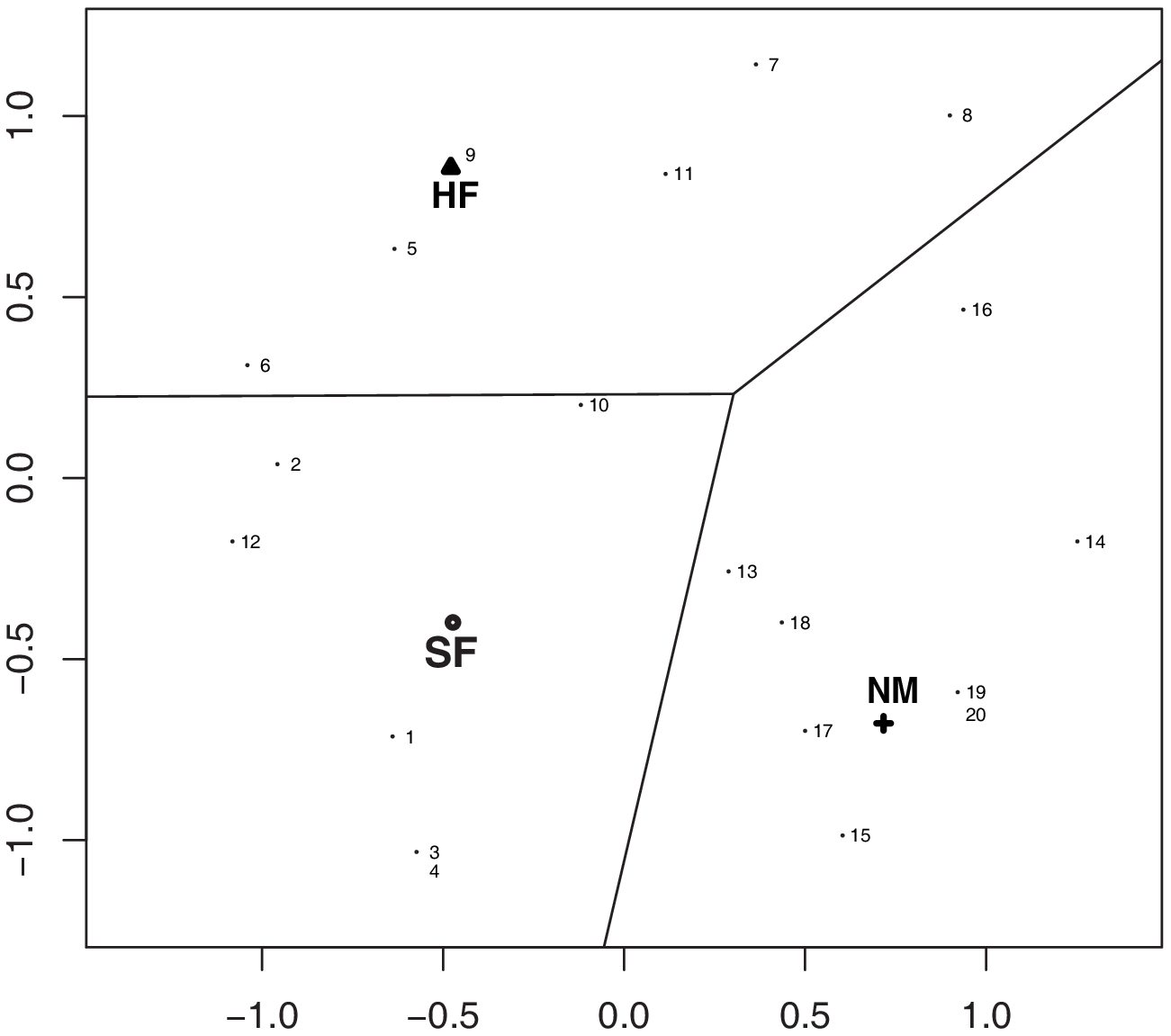}}\\[4pt]
   \subfloat[(c) Grassland use]{\label{fig:c} \includegraphics[width=0.50\textwidth]{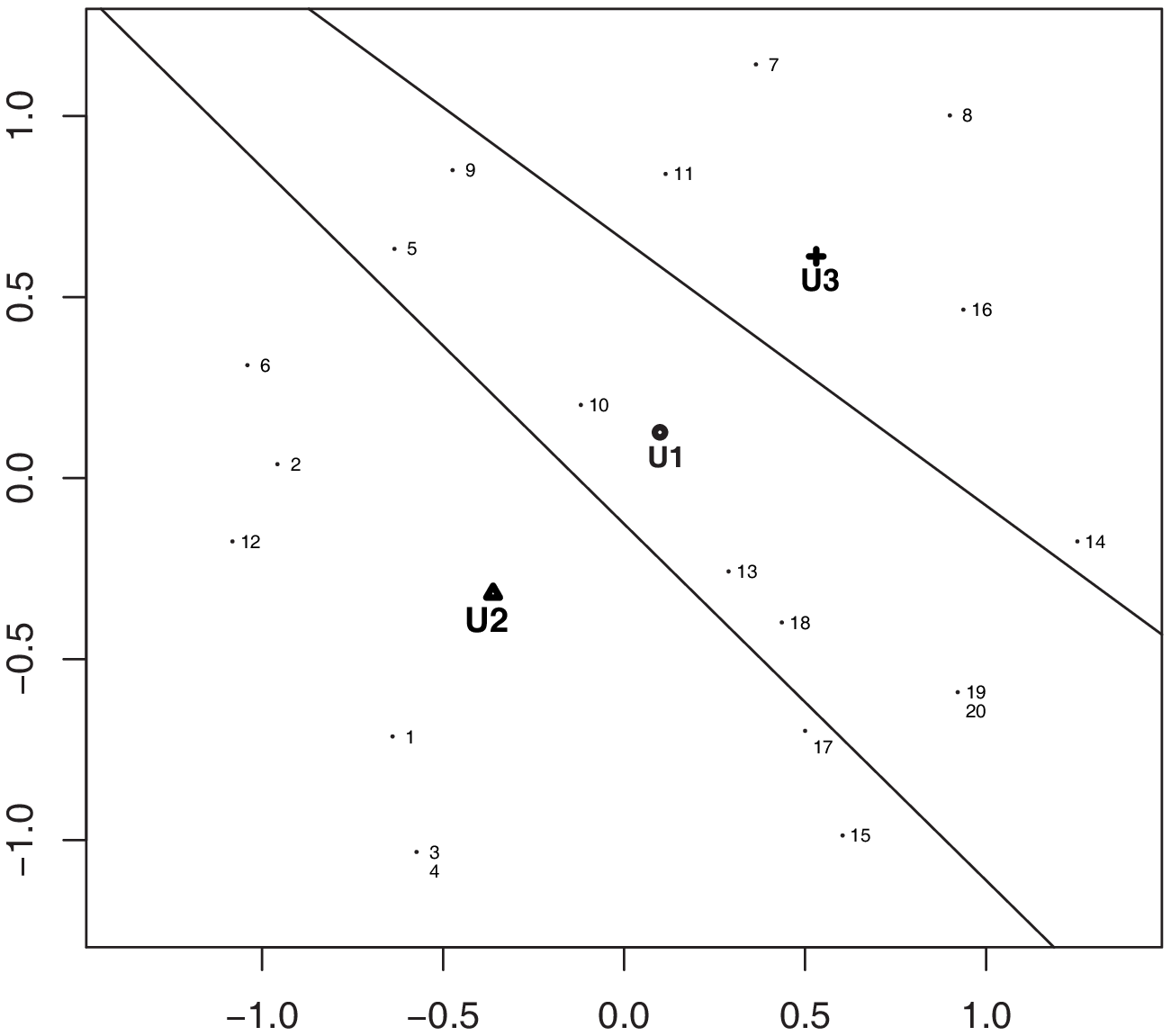}}
   \subfloat[(d) Fertilizer]{\label{fig:d} \includegraphics[width=0.50\textwidth]{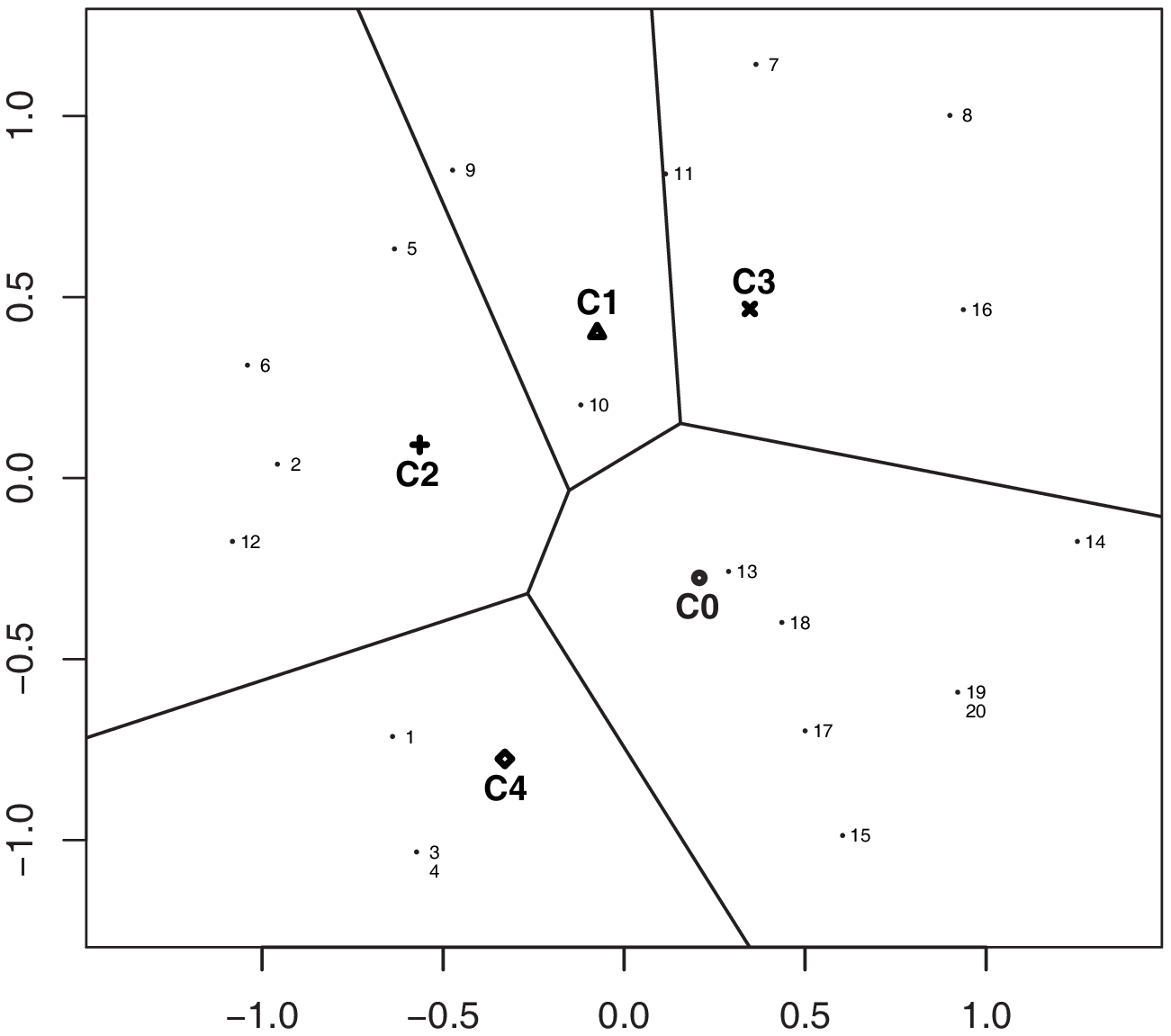}}\\[4pt]
   \caption{The prediction regions for each of the four variables, as given by MLB.}
   \label{predregion}     %% Etiqueta para la figura entera
\end{figure}

The prediction regions obtained from the proposed algorithm together with the category points associated to them, are shown in figure~\ref{predregion}.

\begin{figure}[!htb]
  \centering
    \includegraphics[width=0.7 \textwidth]{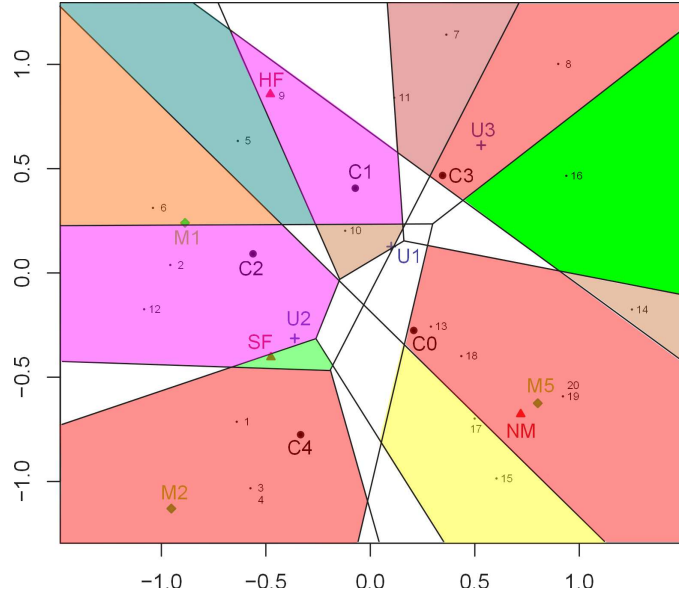}\\[3pt]
  \caption{Four tesselations superimposed.}
  \label{Gower96EMMixed}
\end{figure}

\begin{figure}[!htb]
  \centering
    \includegraphics[width=0.8 \textwidth]{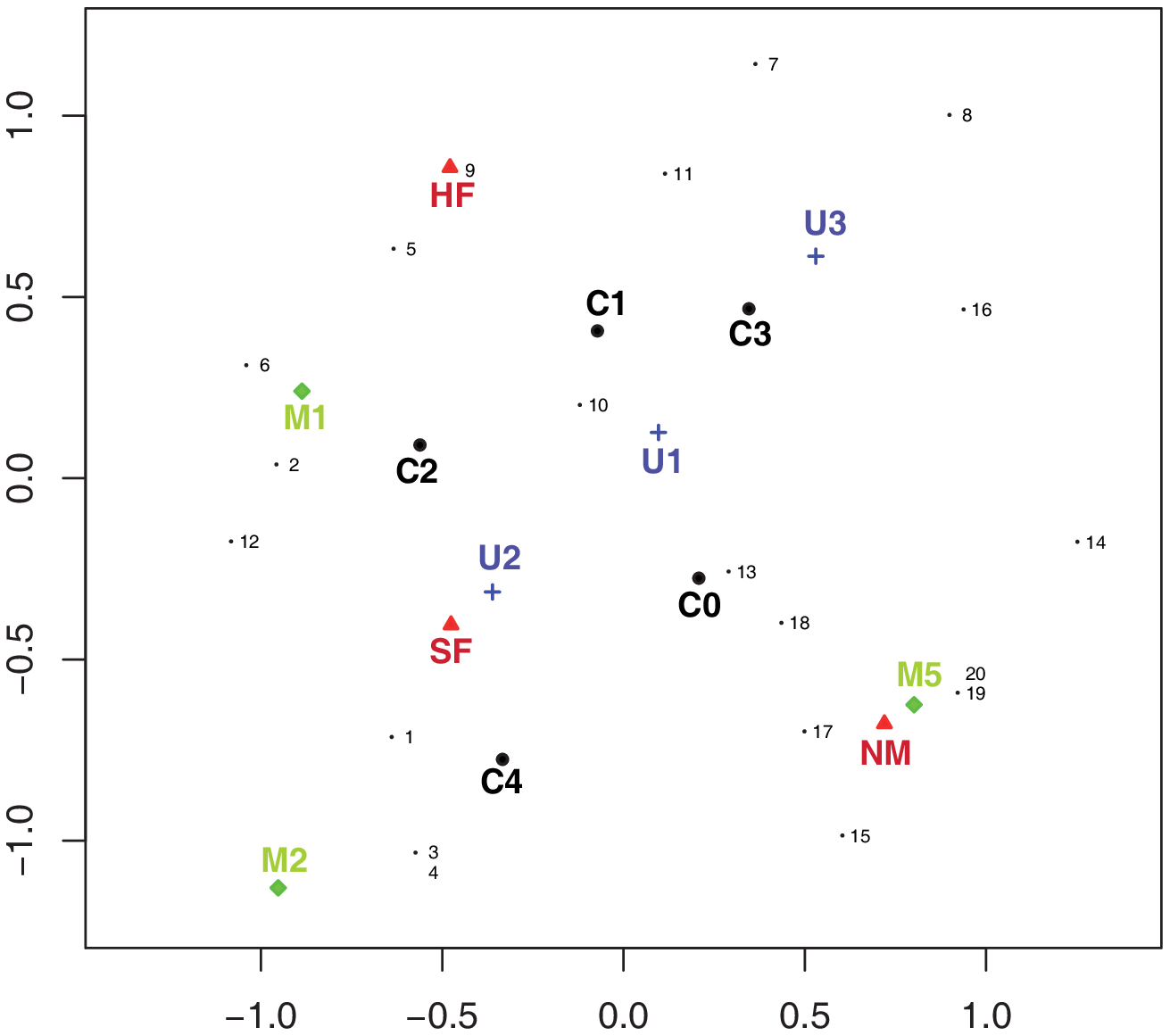}\\[3pt]
  \caption{Two-dimensional BLM of the categorical variables shown in Table~\ref{datafarm}.}
  \label{Gower96EM}
\end{figure}

\begin{table}[!htb]
\begin{flushleft}
\caption{Predictions for the categorical variables for table~\ref{datafarm} given by a two-dimensional approximation. T denotes the true value given in table~\ref{datafarm}, and MCA, Mirt, OP are the predictions using the row coordinates estimated by MCA, Mirt and Alternated Method(AM).}\label{predictions}
\renewcommand{\multirowsetup}{\centering}
\begin{tabular}{|r|l|l|l|l|l|l|l|l|l|l|l|l|l|l|l|l|}
\hline \multirow{2}{0.7cm}{\textbf{Farm}}
 & \multicolumn{4}{p{2cm}|}{\centering \textbf{Moisture}}
 & \multicolumn{4}{p{2cm}|}{\centering \textbf{Management}}
 & \multicolumn{4}{p{2cm}|}{\centering\textbf{Grassland}} 
 & \multicolumn{4}{p{2cm}|}{\centering \textbf{Manuring}}
\tabularnewline \cline{2-17}
& \multicolumn{1}{p{0.15cm}|}{\centering \scriptsize\textbf{T}}
& \multicolumn{1}{p{0.65cm}|}{ \scriptsize\textbf{MCA}}
& \multicolumn{1}{p{0.5cm}|}{\centering \scriptsize\textbf{Mirt}}
& \multicolumn{1}{p{0.40cm}|}{\centering \scriptsize\textbf{AM}}
& \multicolumn{1}{p{0.15cm}|}{\centering \scriptsize\textbf{T}}
& \multicolumn{1}{p{0.65cm}|}{\centering \scriptsize\textbf{MCA}}
& \multicolumn{1}{p{0.5cm}|}{\centering \scriptsize\textbf{Mirt}}
& \multicolumn{1}{p{0.40cm}|}{\centering \scriptsize\textbf{AM}}
& \multicolumn{1}{p{0.15cm}|}{\centering \scriptsize\textbf{T}}
& \multicolumn{1}{p{0.65cm}|}{\centering \scriptsize\textbf{MCA}}
& \multicolumn{1}{p{0.5cm}|}{\centering \scriptsize\textbf{Mirt}}
& \multicolumn{1}{p{0.40cm}|}{\centering \scriptsize\textbf{AM}}
& \multicolumn{1}{p{0.15cm}|}{\centering \scriptsize\textbf{T}}
& \multicolumn{1}{p{0.65cm}|}{\centering \scriptsize\textbf{MCA}}
& \multicolumn{1}{p{0.5cm}|}{\centering \scriptsize\textbf{Mirt}}
& \multicolumn{1}{p{0.40cm}|}{\centering \scriptsize\textbf{AM}}
\tabularnewline \hline
\centering{1}&1&2*&2*&2*&1&1&1&1&2&2&2&2&4&4&4&4\\
\centering{2}&1&1&5*&1&2&3*&3*&1*&2&2&1*&2&2&2&0*&2\\
\centering{3}&2&2&2&2&1&1&1&1&2&2&2&2&4&4&4&4\\
\centering{4}&2&2&2&2&1&1&1&1&2&2&2&2&4&4&4&4\\
\centering{5}&1&1&5*&1&3&3&3&3&1&3*&1&1&2&1*&0*&2\\
\centering{6}&1&1&5*&1&3&3&3&3&2&2&1*&2&2&2&0*&2\\
\centering{7}&1&1&1&1&3&3&3&3&3&1*&3&3&3&1*&3&3\\
\centering{8}&5&1*&1*&5&3&3&3&3&3&1*&3&3&3&3&3&3\\
\centering{9}&4&1*&1*&1*&3&3&1*&3&1&3*&2*&1&1&1&3*&1\\
\centering{10}&2&1*&5*&1*&2&3*&4*&1*&1&1&1&1&1&1&0*&1\\
\centering{11}&1&1&5*&1&2&3*&3*&3*&3&3&3&3&1&1&2*&3*\\
\centering{12}&3&1*&5*&1*&1&1&3*&1&2&2&3*&2&2&2&3*&2\\
\centering{13}&5&2*&1*&5&1&1&1&4*&2&2&2&1*&3&4*&3&0*\\
\centering{14}&5&5&5&5&4&4&4&4&3&1*&1*&3&0&0&0&0\\
\centering{15}&5&5&5&5&4&4&4&4&2&1*&1*&2&0&0&0&0\\
\centering{16}&5&5&1*&5&1&1&1&4*&3&2*&3&3&3&3&3&3\\
\centering{17}&2&5*&5*&5*&4&4&4&4&1&1&1&1&0&0&0&0\\
\centering{18}&1&5*&5*&5*&4&4&4&4&1&1&1&1&0&0&0&0\\
\centering{19}&5&5&5&5&4&4&4&4&1&1&1&1&0&0&0&0\\
\centering{20}&5&5&5&5&4&4&4&4&1&1&1&1&0&0&0&0\\
\hline
\centering\tiny{Errors}&0&8&13&6&0&3&5&5&0&7&6&1&0&3&7&2\\
\hline
\end{tabular}
\end{flushleft}
\end{table} 

The four graphs could be superimposed although the resulting image would be almost unreadable (figure~\ref{Gower96EMMixed}) even with only four variables; with more variables the interpretation would be very complicated. The proposed procedure for obtaining a set of category points for each variable allows for a much simpler and easy to interpret representation. The final result is shown in figure~\ref{Gower96EM}. We can see that farms having a ``nature management'' (NM) are at the areas with higher moisture (M5), zero fertilizer (CO) and hay production (U1). Farms with ``scientific management'' (SF) are at the region with moisture M1 and M2, high values of fertilizer (C4) and intermediate grassland use (U2). Hobby farms (HF) are associated to dry places (M1), low use of fertilizer (C1) and a tendency toward U3. Farms of type BF are hidden on the prediction model because the probability of that category is never higher than the rest.

In order to compare the proposed method with MCA as in \cite{Gower-96}, and some alternatives for estimation described here, we have estimated the model parameters using our modification of the EM algorithm and the mirt package \citep{Chalmers2012} with an additional multinomial logistic regression. The prediction regions obtained for our method produce 14 incorrect classifications against the 21 obtained by MCA and the 31 by mirt (see table~\ref{predictions}). The table shows also true and predicted categories for all the data matrix. There are no hidden categories for variable ``Manuring'' but for ``Moisture'' and ``Management'', categories M4 and BF, respectively, are hidden. The last value is present in farmers 2, 10 and 11 and none of the methods is able to predict it correctly.

If we analyse the combined prediction regions for all the variables with  EM parameter estimation, we can observe in figure~\ref{Gower96EMMixed} that there are 28 separate convex regions. Except region containing farms 13, 18, 19 and 20, most of the regions are small and have less points inside, emphasizing the richness of the technique for interpreting data.  In the study described by \cite{Gower-96}, there were 16 different regions for MCA but only three were clearly populated, so we obtain a finer classification of the farms.

\section{Conclusions and discussion}\label{conc}

In the preceding sections we have proposed a biplot method for nominal data in which the individual are represented as points in a low-dimensional subspace and the variables are represented as ``prediction regions'' or ``category points'' for the categories of each variable. Prediction regions are convex polygons that divide the representation space into as many regions as categories of the variable, except if there is some hidden category, and then define a tessellation of the space that, conveniently approximated by a Voronoi diagram, provides a set of generators that can be considered as category points. The proposed representation is interpreted in terms of distances in the sense that the category predicted for each individual is defined by the closest category points.
Although not described here in detail, linear biplots for the log odds of each category with the baseline.

A simple adaptation of an EM-algorithm is proposed for estimation of model parameters. The usual alternated EM algorithm is modified to include penalized ridge estimation of the logistic model parameters in order to avoid  the problems produced by the separation that makes the estimators undefined. Other penalized methods are the lasso for logistic regression \citep{Meier2008}, the Firth method \citep{Firth1993} applied to multinomial models by \cite{Bull2002}. The estimators obtained from the package mirt \citep{Chalmers2012} can also be used as a start point to construct the biplot, using the factor scores but with an additional step to refit the nominal logistic model for the variable parameters. This is so because mirt is designed for Item response theory, the scores are always calculated with an additional rotation but the parameters seems not to be rotated consequently. In some examples we have tried the numerical values are strange probably due to the fact that mirt does not take into account the separation problem. Both, our alternated method and mirt perform better when the number of individuals are much higher than the number of variables but there are many practical problems in which this is not so, for example, trying to classify a set of individuals with the genotypes resulting from thousands of single nucleotide polymorphisms \citep{Demeyetal}. For those cases it is probably more efficient to estimate the individual markers by principal coordinates of the matrix  $\bf{G}$ of indicators defined previously and then fitting the nominal models on the coordinates. This is not a maximum likelihood solution but it is a good approximation when the other methods are unstable.
The main advantage of using maximum likelihood is that it is possible to perform hypothesis testing to compare different models, for example to select the number of dimensions to retain. The proposed method share the characteristics of ``formal'' models as item response theory or latent traits and ``descriptive'' models as MCA, could even be considered also as a graphical representation of the formal model.
It has to be noted that the performance of the algorithm for approximation and inversion of the tessellation crucially depends on the goodness of fit of the nominal regression. Only variables with a reasonable fit should be represented on the graph.

 \section{Software Note}\label{soft}
 An R package containing the procedures described by this paper has been developed by the authors \citep{NLB}.

\end{document}